\documentclass[aps,prb,twocolumn,groupaddress,showkeys]{revtex4}
\usepackage{graphics}
\usepackage{graphicx}
\usepackage[centertags]{amsmath}

\begin{document}

\title{Influence of Morphology on the Optical Properties of Metal Nanoparticles}

 \author{A. L. Gonz\'alez}
 \affiliation{Instituto de F\'{\i}sica, Universidad Nacional Aut\'onoma
de M\'exico, Apartado Postal 20-364, D.F. 01000,  M\'exico.}

\author{Cecilia Noguez}
\email[Corresponding author. Email:]{cecilia@fisica.unam.mx}
\affiliation{Instituto de F\'{\i}sica, Universidad Nacional Aut\'onoma
 de M\'exico, Apartado Postal 20-364, D.F. 01000,  M\'exico.}

   \date{Submitted: May 1, 2006. Accepted: July 10, 2006}

\begin{abstract}

The influence of morphology on the optical properties of silver nanoparticles is studied.  A general relationship between the surface plasmon resonances and the morphology of each nanoparticle is established. The optical response is investigated for cubes and decahedrons with different truncations.  We found that polyhedral nanoparticles composed with less faces show  more surface plasmon resonances than spherical-like ones. It is also observed that the vertices of the nanoparticles play an important role in the optical response, because the sharpener they become, the greater the number of resonances. For all the nanoparticles, a main resonance with a dipolar character was identified as well as other secondary resonances of less intensity. It is also found that as the nanoparticle becomes more symmetric, the main resonance is always blue shifted. 
\end{abstract}

\maketitle


\section{Introduction}\label{intro}

New synthesis methods have been developed to fabricate nanoparticles (NPs) with specific size and shape that  in turn have enabled us to control optical properties to reveal new aspects of their underlying science and to tailor them for clearly defined applications. For instance, the optical response of nanoparticles is now being investigated for their potential in optics, magneto-optic, photonics, as a nanoengineered substrate on which the Surface Enhanced Raman Scattering (SERS) response can be precisely controlled and optimized, and for chemical and biosensing applications, such as optical addressable diagnostic methods, devices, and therapies  based on the plasmonic response of metallic NPs.~\cite{ozbay,hibbins,barnes} These advances, that allow metals to be structured and characterized on the nanometer scale, have renewed the interest in optical properties from physicists, chemists and materials scientists to biologists. 

The optical properties of metal nanoparticles (NPs) can be tuned by controlling their size and shape. Indeed, metallic NPs support surface plasmon  resonances, which are highly dependent on geometry and environment, providing a starting point for  emerging  research fields like plasmonics.~\cite{ozbay} Surface plasmons are collective excitations of the electrons at the interface between a conductor and an insulator, and are described by evanescent electromagnetic waves which are not necessarily located at the interface.  Surface plasmon resonances (SPR) appear in a number of different phenomena, including the optical response of materials at different scales, and the Casimir and van der Waals forces between macroscopic bodies. 

Besides to the technological implications, the optical  signature of NPs  can be also used as a tool of characterization.  Optical techniques are non-destructive and, with a proper implementation, they can be used to perform \textit{in situ} and in real time measurements,  providing statistical properties of the whole sample. These attributes are important  because the properties of nanoparticles depend on the environment.~\cite{noguez,kelly} When growth and characterization are  made in different ambient conditions, this can be an additional uncontrollable variable for their interpretation.  Thus, optical spectroscopies can be used as complementary tools of structural characterization techniques like Atomic Force Microscopy (AFM), Scanning Tunneling Microscopy (STM), Transmission Electron Microscopy (TEM), etc., which provide the image of a small piece of the sample, giving information about local properties and characterizing a few NPs at a time. 

 The interesting observation that NPs support SPR that can be tuned by controlling their size and shape, offers a starting point for nanotechnology.~\cite{noguez,kelly,sosa,gonzalez}  In this article, we give some insights of the SPR as a function of the morphology for silver NPs of different polyhedral shapes. In the case of silver, many results indicate the presence of icosahedra, and decahedral shapes as well as other related morphologies like cubes and truncated cubes.~\cite{gonzalez} A very similar pattern is found in 
gold and copper.~\cite{baletto,baletto05}


\section{Formalism}

\begin{figure}[htbp]
\begin{center} 
\includegraphics[width=  0.5\textwidth]{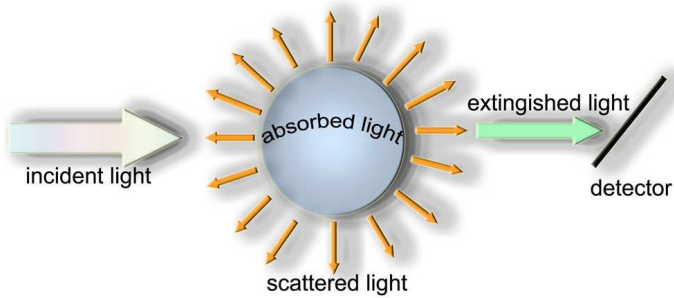}
\caption{Schematic model of light extinction due to scattering and absorption effects.} \label{fig1}
\end{center}
\end{figure}

When a particle is under the action of an electromagnetic (EM) field, its electrons start to oscillate, transforming energy from the incident EM wave into, for example, thermal energy in a so-called absorption process. The electrons can also be accelerated and then, they can radiate energy in a so-called scattering process. The sum of both effects, absorption and scattering, is known as the extinction of light, see Fig.~\ref{fig1}. In this work, we consider NPs, which are large enough to employ the classical EM theory. However, they are small enough to observe the dependence of the optical properties with its size and shape. This means that the inhomogeneities of the particle are much smaller as compared to the wavelength of the incident  EM field, such that, each point of the nanoparticle can be described in terms of its macroscopic dielectric function, which depends on the frequency only. Here, we restrict ourselves to the elastic or coherent case, where the frequency of the absorbed and scattered light is the same as the frequency of the incident light. 

When the size of a homogeneous particle is much smaller than the wavelength of the incident light,  the NPs feel a field spatially constant, but with a time dependent phase. This is known as the quasi-static approximation, which is characterized by keeping the time but not the spatial dependence of the EM field. The electric field causes the conduction electrons to oscillate coherently, displacing the electron cloud from the nuclei. The positive charges are assumed to be immobile, while the negative charges are allowed to move under the influence of the incident field. The long range correlation of electrons caused by Coulomb forces, and the attraction between positive and negative charges results in a restoring force that changes the oscillation frequency of the electron cloud with respect to the positive background.  In metallic particles, the collective motion of electrons produces surface modes, whose number, position, and width are determined by the particle shape and variations of the dielectric function. Since the main effect producing the restoring force is the surface polarization, the proper resonances depend on the NP shape. For example, ellipsoids with three different axes have  three different oscillation frequencies, depending on the relative length of the axes, such that, the optical response of the NP is sensitive to the choice of the light polarization.

For nanoparticles of less than $ 10$~nm, radiation processes are negligible, then the particle only absorbs energy through the excitation of surface plasmon resonances.~\cite{noguez} Altering the surface, by modifying the size, shape, and/or environment of the conductor, the properties of SPR can be tailored.  SPR of small particles can be studied in terms of the strength of the coupling to the applied field of the optically active electromagnetic surface modes of the system. These modes only depend on the morphology of the particle, and not on its material properties.~\cite{roman}

The optical response of a NP characterized by a dielectric function $\epsilon(\omega)$, can be obtained by finding the solution of the Maxwell's equations. In 1908, Gustav Mie found the exact answer for a homogeneous spherical particle of arbitrary size.~\cite{mie} However, exact solutions of the Maxwell's equations for particles with other arbitrary shapes are not straight forward, and only approximations are possible. Because of the complexity of the systems being studied here, efficient computational methods capable of treating large size systems are essential. In the last few
years, several numerical methods have been developed to determine the optical properties of non-spherical particles, such as the Discrete Dipole Approximation, T-matrix, Spectral Representation, Finite Difference methods, etc.~\cite{mish} In this work, we employ the Discrete Dipole Approximation.

\subsection{Discrete Dipole Approximation}

The Discrete Dipole Approximation (DDA) is a well-suited technique for studying scattering and absorption of electromagnetic radiation by particles
with sizes of the order or less of the wavelength of the incident light. DDA
has been applied to a broad range of problems, including interstellar dust
grains, ice crystals in the atmosphere, interplanetary dust, human blood
cells, surface features of semiconductors, metal nanoparticles and their
aggregates, and more. DDA was first introduced by Purcell and
Pennypacker,~ \cite{purcell} and has been subjected to several improvements,
in particular those made by Draine, and collaborators.~\cite{draine1,draine2,draine3} Below,
we briefly describe the main characteristics of DDA and its numerical
implementation: the DDSCAT code.~\cite{DDSCAT} For a more complete description of DDA and
DDSCAT, the reader can consult Refs.~[12--16].

\begin{figure}[htbp]
\begin{center} 
\includegraphics[width=  0.5\textwidth]{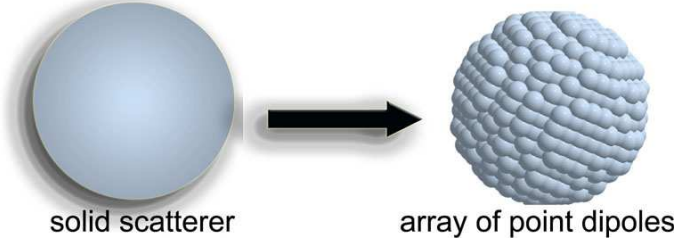}
\caption{DDA approximates a solid scatterer by an array of polarizable point dipoles that mimics the morphology.} \label{fig2}
\end{center}
\end{figure}

\textsc{DDSCAT} builds up a solid object using an array of $N$ polarizable entities  located  in a periodic lattice which resembles the shape and size of the particle under study, see Fig.~\ref{fig2}. These polarizable entities are located at the positions $\vec{r}_i$ with $i = 1, 2, \dots N$.  \textsc{DDSCAT}  assigns to each entity a dipole moment given as
\begin{equation} \label{dda1}
\vec {p}_i=\overleftrightarrow{\alpha}_i \cdot \vec {E}_{i,\text{loc}} \,,
\end{equation}
where $\overleftrightarrow{\alpha}_i $ is the dipolar polarizability of the entity at $\vec{r}_i$, and $\vec {E}_{i,\text{loc}}$  is the total electric field acting on the $i$-th dipole, also called the local field.  The discretization of the particle is a good approximation when the distance  $d$ between adjacent polarizable entities is much smaller than the wavelength $\lambda$ of the incident electromagnetic field. Once $\vec{r}_i$, and $\overleftrightarrow{\alpha}_i $ are given,  and the condition $d/\lambda\ll 1$ is fulfilled,  it is possible to predict the light absorption and scattering by free and embedded  particles. In our case, the silver nanoparticles of interest  are suspended  in a solvent in such a way that both solvent and particles constitute a dilute colloidal suspension.

The local field due to an array of point dipoles under an applied electromagnetic field is given as 
\begin{equation} \label{Elocal}
\vec{E}_{i,\text{loc}}=\vec{E}_{i,\text{app}}+\vec {E}_{i,\text{ind}} \, ,
\end{equation}
where $\vec{E}_{i,\text{app}}$ is the applied field, and $\vec {E}_{i,\text{ind}}$ is the  induced field  acting on the $i$-th  entity due to the radiation of all the others $N-1 $ dipoles that compose the nanoparticle. 

Let us consider the applied field  as a monochromatic plane wave \[
\vec E_{i,\text{app}}=\vec E_0\exp(i\vec k\cdot\vec r-\omega t)
,\]  where $ E_0$ denotes the magnitude of the incident electric field, $\vec{k}$ \ the wave vector, $\omega$ the frequency and $t$ means time. On the other hand, the induced field is given by
\begin{equation} \label{Eind}
\vec{E}_{i,\text{ind}}=\sum_{j=1}^N {'} \overleftrightarrow{A}_{ij} \cdot \vec{p}_j \, ,
\end{equation}
where the symbol $(')$ means $i \neq j$ and $\overleftrightarrow{A}_{ij}$ is  the matrix that couples the electromagnetic interaction among dipoles. This interaction matrix is expressed as
\begin{eqnarray} \label{aij}
\overleftrightarrow{A}_{ij} \cdot \vec{p}_j&=&\frac{e^{ikr_{ij}}}{r_{ij}^3} \Bigg \lbrace k^2 \vec{r}_{ij} \times (\vec{r}_{ij} \times \vec{p}_j) \\ \nonumber
&+& \frac{(1-ikr_{ij})}{r_{ij}^2} \left[ r_{ij}^2 \vec{p}_j- 3\vec{r}_{ij}(\vec{r}_{ij} \cdot \vec{p}_j) \right] \Bigg \rbrace  \, ,
\end{eqnarray}
where $\vec{r}_{ij}$ is a vector from the position of dipole \emph{i}-th to dipole \emph{j}-th, $r_{ij}$ denotes its magnitude, and $k =|\vec{k}|$. 

Substituting  the local field from Eqs.~(\ref{Elocal}) and (\ref{Eind}) into the expression for the dipolar moment in Eq.~(\ref{dda1}), we obtain a system of $3N$ complex coupled equations:
\begin{equation} \label{dipolos}
\vec{p}_i= \overleftrightarrow{\alpha}_i \cdot \left( \vec{E}_{i,\text{inc}}+{\sum_{j=1}^{N}} {'} \overleftrightarrow{A}_{ij} \cdot  \vec{p}_j \right) \, .
\end{equation}
From the above expression, we can calculate the set of dipole moments that mimic the optical response of the particle; and 
once we know  each $\vec{p}_i $, it  is possible to obtain the light extinction and absorption cross sections using the following expressions:~\cite{purcell}
\begin{equation} \label{seccion eficaz}
C_{\text{ext}}=\frac{4\pi k}{|\vec{E}_0|^2}\sum_{j=1}^N\mbox{Im}\{\vec{E}_{j,\text{inc}}\cdot \vec{p}_j^* \},
\end{equation}
\begin{equation}
C_{\text{abs}}=\frac{4\pi k}{|\vec{E}_0|^2}\sum_{j=1}^N \big \lbrace \mbox{Im}[\vec{p}_j \cdot(\alpha_j^{-1})^* \vec{p}_j^*] -\frac{2}{3}k^3 |\vec{p}_j|^2 \big \rbrace \, ,
\end{equation}
where ($\ast$) means complex conjugate. The scattering cross section $C_{\text{sca}}$ is defined as the difference between extinction and absorption cross sections, $C_{\text{sca}} = C_{\text{ext}} - C_{\text{abs}}$.

\textsc{DDSCAT}~\cite{DDSCAT} creates a cubic lattice array of dipoles and assigns  to each one of they a polarizability given by the Lattice Dispersion Relation (LDR):~\cite{draine2}
\begin{equation}
\alpha^{\rm LDR}=\frac{\alpha^{\rm CM}}{1 + \alpha^{\rm CM}\left[b_1+ b_2 {\epsilon}+ b_3 S {\epsilon} \right](k^2/d)},
\end{equation}
where $\epsilon$ is the macroscopic dielectric function of the particle; $S$, $b_1$, $b_2$, and $b_3$ are the coefficients of the expansion to third order in $k$ to incorporate radiation effects, and $\alpha^{\rm CM}$ is the polarizability given by the well known Clausius-Mossotti relation:~\cite{Electro}
\begin{equation}
\epsilon-1=\frac{4 \pi n \alpha^{\rm CM}}{1-4 \pi n \alpha^{\rm CM}/3}   \, .
\end{equation}
Here, we have assumed that the polarizability  is isotropic and is the same for all the entities, $\overleftrightarrow{\alpha_i} = \alpha^{\textrm{LDR}} $.  A key factor in determining the level of accuracy that can be reached for a given number of dipoles is the prescription for assigning dipole polarizabilities.~\cite{draine2} Besides the finite wavelength, it is also important to include the influence of the geometry of the particle has to be considered in the dipole polarizabilities.~\cite{rah} 

It is also important to select a ``realistic'' dielectric function, which better resembles the material properties of the particle at the nanometer scale.   
As starting point, we can employ dielectric functions measured experimentally for bulk metals, $\epsilon_{\text{bulk}}(\omega)$, and then incorporate the main effects at the appropriate scale. The experimental dielectric function has contributions due to \emph{interband} (inter) and  \emph{intraband} (intra) electron transitions, and assuming that both contributions are additive, we have
\begin{equation}
\epsilon_{\text{bulk}} (\omega)=  \epsilon_{\text{inter}}(\omega) + \epsilon_{\text{intra}}(\omega)\, .
\end{equation}
\emph{Interband} contributions are due to electron transitions from occupied to empty bulk bands separated by an energy gap.  The electrons are bound for a restoring force given by the energy difference between ground and excited electron states, usually at the ultraviolet (UV) region for metals.~\cite{mie} 
 \emph{Intraband} contributions come from electron transitions at the Fermi level in incompletely filled bands, or an otherwise when  a filled band overlaps in energy with an empty band. These transitions also provide an absorption mechanism but at lower energies, in metals from infrared (IR) to visible light. 
Electrons at the Fermi level in metals can be excited by photons with very small energies, such that, we say that electrons are ``free'', and their contribution to $\epsilon_{\text{bulk}}(\omega)$, can be approximated by the Drude model of free electrons:~\cite{mie}
\begin{equation}
\epsilon_{\text{intra}}(\omega) = 1 -  \frac{\omega_p^2}{\omega(\omega + i/\tau)} \, . \label{drude}
\end{equation}
Here  $\omega_p = 4 \pi \rho e^2/m$ is the plasma frequency with $\rho$ the number of electrons per volume unit, $e$ the electron charge, $m$ the electron mass, and  $1/\tau$ a damping constant due to the dispersion of the electrons.  At low temperatures, collision times $\tau$ are determined by impurities and imperfections in the lattice, while at room temperatures they are dispersed by the ions of the system. 

Here, we are interested in nanoparticles whose sizes are smaller than $10$~nm, such that, physical phenomena associate with radiation effects, like scattering and radiation damping, are negligible, i.e., $C_{\text{sca}}\approx 0$, such that, $C_{\text{ext}} = C_{\text{abs}}$.~\cite{noguez} However, we have to consider  that the conduction electrons suffer an additional damping effect due to surface dispersion or finite size. In such case,  we have to make an additional adaptation to $\epsilon_{\text{bulk}}(\omega)$,  including an extra damping term $\tau(a)$, which depends on the size $a$ of the particle. This surface dispersion is present when the mean free path of the ``free'' electrons is comparable or larger than the dimension of the particle, such that, the electrons are scattered by the surface rather than the ions. The smaller the particle, the more important are the surface dispersion effects. The surface dispersion  not only depend on the particle's size, but also on its shape.~\cite{coronado}  

To include surface dispersion we need modify the free electron or  \emph{intraband} contributions by changing the damping term.  Assuming that the susceptibilities are additive, we can obtain the effect of the bound charges  by subtracting the free electron contribution, $\epsilon_{\text{intra}} (\omega)$, from the bulk dielectric function. The  free electron or  \emph{intraband} contributions are obtained using the Drude model and the theoretical values of $\omega_p$. Then, we include the surface dispersion by adding the extra damping term $\tau(a)$ to the Drude model. 
Finally, we obtain a dielectric function, which also depends on the NP's size, and includes the contributions of (i) the free electrons, (ii) surface damping, and (iii) \emph{interband} or bound electrons effects, and is given by
\begin{equation}
  \epsilon (\omega,a)
  = \epsilon_{\text{bulk}} (\omega)- \epsilon_{\text{intra}} (\omega)+\Bigg\{1 -\frac{\omega_p^2}{\omega(\omega + i/\tau +i/\tau(a))} \Bigg\}  . \label{fdiele}
\end{equation}
 In this work, we will consider for all the cases the surface dispersion of a sphere of radius $a$ is given by   $1/\tau(a)=v_f/a$,~\cite{kreibig}  where  $v_f$ is the Fermi velocity of the electron cloud. We will show  that surface dispersion effects do not change the location of the surface modes, but they affect the coupling of the proper modes with the applied field, making wider and less intense.

\section{Results and Discussion}

To understand the influence of morphology on the SPR, we study the extinction efficiency $Q_{\mathrm{ext}}$, for different polyhedral NPs,  defined as the extinction cross section per unit area $A$,  as follows   
\begin{equation}
Q_{\mathrm{ext}}=\frac{C_{\mathrm{ext}}}{A} \, .
\end{equation} 
  We consider silver particles
with an effective volume $4\pi a_{\mathrm{eff}}^{3}/3$, with $a_{\mathrm{eff}
}=$10~nm.  In all cases, the spectra were calculated within DDA using more than $10^5$ dipoles to assure convergence. We employ the dielectric function as measured on bulk silver  by Johnson and Christy,~\cite{johnson} and modified according to Eq.~(\ref{fdiele}). We assume that the  NPs are immersed  in a media with a refraction index $n=1$,  and are well dispersed at a low concentration. In this dilute regime  the interactions between particles are negligible,~\cite{barrera} such that, the absorbance can be modeled as the optical response of one immersed NP times the concentration of particles.~\cite{gonzalez}

\subsection{Cubic morphology}

\begin{figure}[htbp]
\begin{center} 
\includegraphics[width=  0.5\textwidth]{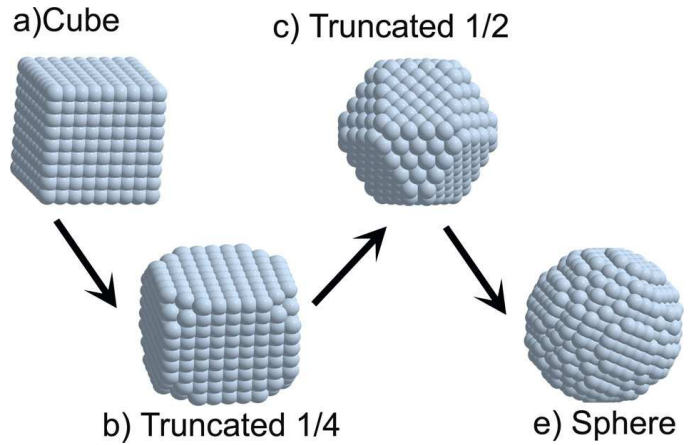}
\caption{Cube nanoparticle, and two different truncated cubes, and the sphere.} \label{fig3}
\end{center}
\end{figure}

We study the extinction efficiency of cubic particles and compare it with those obtained for different truncated cubes and the sphere.  The truncated cubes are obtained by symmetrically edging the eight vertices of the cube by $l \times r$, where $l$ is the length of the cube's side and $0 < r \leq 1/2$. We label the different truncations  with the number $r$. When $r = 1/2$  a cuboctahedron is obtained. Six octagons and eight triangles compose all the truncated cubes, while the cuboctahedron is  composed by six planar  squares and eight triangles. All the truncated cubes have fourteen faces. Finally, if we performed a symmetric truncation of the cube with an infinite number of planes, one could arrive to the sphere, as shown in Fig.~\ref{fig3}.

\begin{figure}[htbp]
\begin{center} 
\includegraphics[width=  0.5\textwidth]{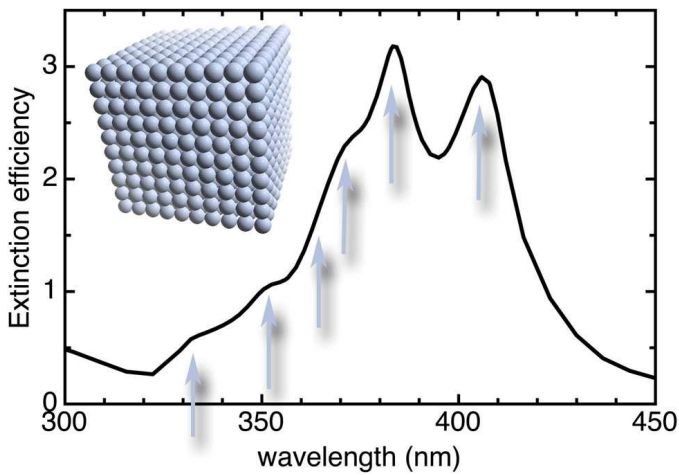}
\caption{Extinction efficiency of a silver cube nanoparticle as a function of the wavelength of the incident light. The main six surface plasmon resonances are indicated by arrows.} \label{fig4}
\end{center}
\end{figure}

In Fig.~\ref{fig4}, we show the extinction efficiency of a silver nanocube. As we will show later, the structure below 335~nm is  independent  of the morphology of the particle,  because the main absorption mechanism at those wavelengths is the \emph{interband} component. At larger wavelengths,  the spectrum is very sensitive to the morphology of each NP. For the cubic NP, the spectrum shows a rich structure of peaks, contrary to the case of the sphere that has a single resonance. These peaks are associated to the SPR inherent to the cubic geometry. Fuchs~\cite{fuchs} found nine SPR for the cubic geometry, where only six of them account for more than the 95~\% of the spectrum, as seen in Fig.~\ref{fig4}. The two most intense SPRs correspond to the dipolar and quadrupolar charge distributions, and are located at 406~nm and 383~nm, respectively.  The other SPRs are at smaller wavelengths ($< 370$~nm), making  wider the spectrum. Let us remind you that the small nanosphere shows a single peak because only a homogeneous arrangement of the charges is possible, giving rise to a dipolar  charge distribution. On the other hand,   small cubes have more resonances  because the charges are not longer able to arrange in a homogeneous distribution, resulting in many different ways beside the dipolar charge distribution, even in the long wavelength approximation.~\cite{fuchs}

\begin{figure}[htbp]
\begin{center} 
\includegraphics[width=  0.5\textwidth]{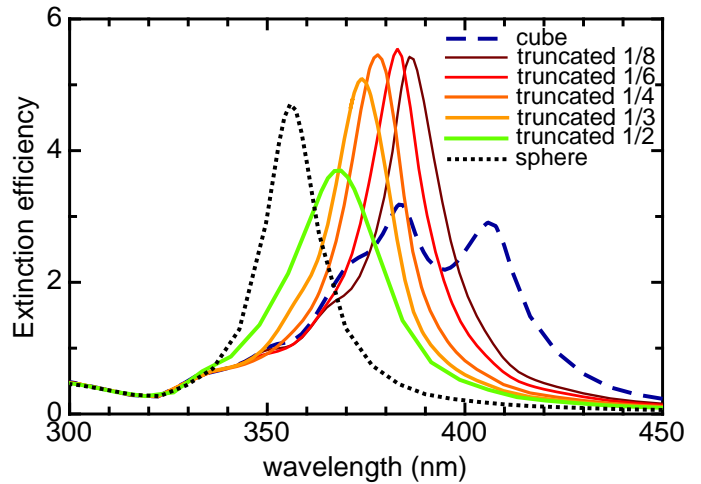}
\caption{Extinction efficiencies as a function of the wavelength of the incident light of a silver cube, different truncated cubes and  a spherical nanoparticle.} \label{fig5}
\end{center}
\end{figure}

In Fig.~\ref{fig5}, the extinction efficiencies of truncated nanocubes with $r$ from $1/8$ to $1/2$ (cuboctahedron) are shown in solid lines. The spectra for  spherical (dotted line) and cubic (dashed line) NPs are also included for comparison. It is observed that even for the smallest truncation of $r=1/8$, the spectrum is very sensitive to the morphology. The dipolar resonance is blue shifted by about 20~nm, and now becomes the most intense resonance. The location of the dipolar and quadrupolar resonances are now very close, such that, only one wide peak is observed around 386~nm, while the structure below 370~nm remains almost identical to the spectrum of the cube. The same trend is observed for larger truncations, and from  Fig.~\ref{fig5}  we find that as the length-size of the truncation increases: 
 (i) the main resonance is always blue shifted,  (ii) the peaks  at smaller wavelength are closer to the dominant resonance, such that, they  are hidden, and (iii) the width of the main resonance increases.
 For instance, the full width at the half maximum 
(FWHM) of the $1/8$ truncated cube is about 20~nm, while the one for the cuboctahedron is 40~nm. This means that the secondary resonances do not disappear but they are hidden by the dominant  resonance. For comparison, we have included the spectrum of a silver nanosphere of  10~nm of diameter. In this case, the sphere shows a single SPR  located at  356~nm and shows a FWHM of 15~nm.  For icosahedra NPs (not shown here), we have found the main SPR at 363~nm with a FWHM of 25~nm. We can conclude that as the number of faces of the NP increases the energy range of the spectrum becomes smaller, the main resonance is blue shifted, the FWHM decreases, and fewer resonances are observed. Therefore, by obtaining small differences in morphology, it is possible tune SPRs at different wavelengths.

\subsection{Decahedral morphology}

\begin{figure}[htbp]
\begin{center} 
\includegraphics[width=  0.5\textwidth]{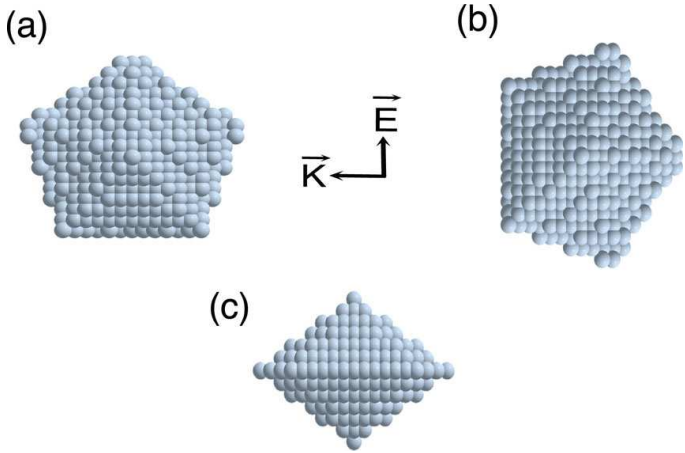}
\caption{Regular decahedral or pentagonal bipyramid nanoparticle and its three different orientations to the incident electromagnetic field. (a) parallel to the pentagon along the vertices,  (b) parallel to the pentagon along the edges, and (c) perpendicular to the pentagon.} \label{fig6}
\end{center}
\end{figure}

Another important morphology present in  metal NPs is the decahedron or pentagonal bipyramid, which is obtained by different  synthesis methods.~\cite{wang,yacaman,kuo,afm,freund}  It is known that metal nanoparticles of a few nanometers in size show different structural motifs depending on their size, composition and energetic conditions.~\cite{baletto,baletto05} The regular decahedron is composed with ten planar triangular faces which resemble two pentagons, as seen in Fig.~\ref{fig6}, where three different orientations are shown. The decahedron is  an asymmetric particle, such that, the optical response depends of the orientation of the incident electromagnetic field. In Fig.~\ref{fig6}, we show the three different orientations of the regular decahedron with respect to the incident electromagnetic field, where in (a)  the electromagnetic field $\vec{E}$, is parallel to  the pentagonal motif and $\vec{E}$ is along the vertices, in (b) $\vec{E}$ is also parallel but is along the edges, while in (c) $\vec{E}$ is perpendicular to the pentagonal motif. 

\begin{figure}[htbp]
\begin{center} 
\includegraphics[width=  0.5\textwidth]{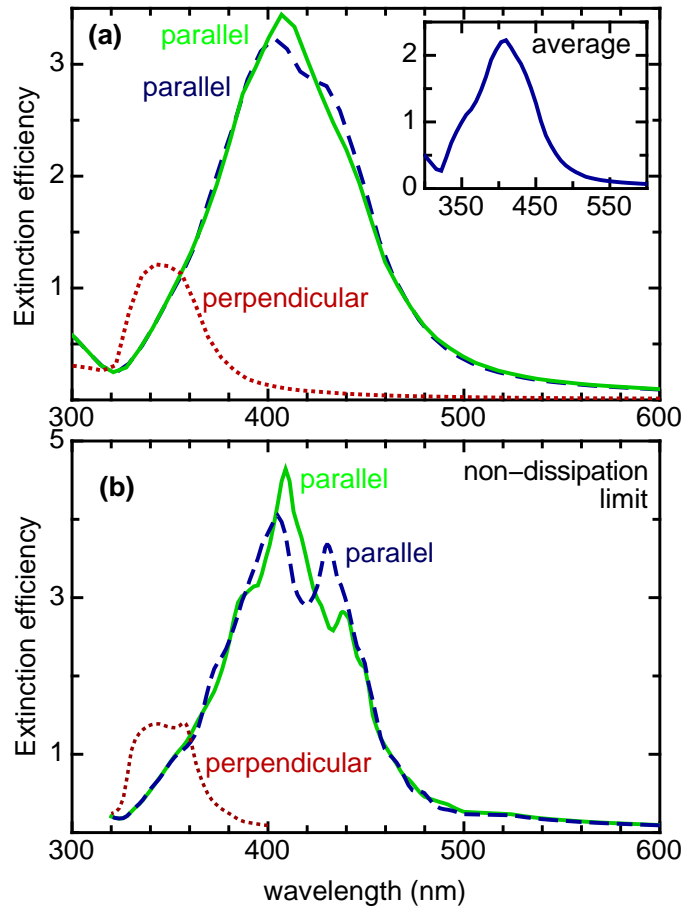}
\caption{(a) Extinction efficiency as a function of the wavelength of the incident light of regular decahedral nanoparticles for different light polarizations. The inset shows  the orientational average. (b) The same but in the non-dissipation limit.} \label{fig7}
\end{center}
\end{figure}

In Fig.~\ref{fig7}(a), the extinction efficiency of the decahedral particle for the three different polarizations, as well as  their average (inset)  are shown. The dashed and solid lines correspond to the case of parallel polarization (a) and (b), respectively, while the dotted line corresponds to the perpendicular polarization (c).   We find a large anisotropy of the extinction when the light incidence is such that the electric field is  parallel and perpendicular to the pentagonal motif.  When the electric field is parallel to the pentagon, the corresponding spectra  are very wide with a FWHM of 90~nm, and a maximum at about 403~nm. On the other hand, when the electric field is perpendicular to the pentagon, the spectrum shows a maximum at about 343~nm, is about three times less intense, and has a FWHM of  45~nm. The spectra for both parallel polarizations are almost identical, except near the maxima, where small differences are observed. These differences would be discussed below. On the other hand, the maxima of the average spectrum is at 410~nm, and the FWHM is about 90~nm. In conclusion, we find that the parallel polarization dominates the average spectrum. The morphology of the decahedral NP shows several SPRs in a wide range of wavelengths. However they are not observed because (a) they are close of each other, such that, the most intense hides the others and/or (b) dissipation effects make wider the resonances, and the detailed structure is ``washed out''.~\cite{noguez} 

We have already mentioned that the location of the resonant frequencies of the proper modes of the system is not immediate, because it requires taking the non-dissipation limit. Here, to find the location of each one of the SPRs for each morphology, we have considered that the constants $\tau$ and $\tau(a)$ in Eq.~(\ref{fdiele}) tend to infinity to achieve this limit. In Fig.~\ref{fig7}(b), we show  the  extinction efficiency corresponding to the three different polarizations described above, but taking the non-dissipation limit. Here, we clearly observe that the peaks become more pronounced, as well as new peaks appear, such that,  we can distinguish the different resonances that compose the spectra. For instance, the optical response for the perpendicular polarization is composed of at least two different SPRs at 343 and 357~nm of about the same magnitude. 
For the spectra corresponding to parallel polarized light, we find several SPRs. In the case when the electric field is along the vertices of the pentagon, as shown in Fig.~\ref{fig6}(a), we identify at least eight different resonances. The first one is at 353~nm that gives rise to a small shoulder, as well as other resonances at 372, 437, 447, 464, 478 and 492~nm. The SPRs at 437 and 447 are also present when the electric field is parallel and along the edges of the pentagons. For the case along the vertices (dashed line), the main resonances are at 405~nm and 430~nm and their intensity are very similar. While for the polarization along the edges, the main SPRs are red shifted to 409~nm and 447~nm, respectively. The first one is almost twice as intense as the other. These dissimilarities in location and intensity of the resonances are responsible for the small difference observed in the spectra. However, the main difference is between light polarized parallel and perpendicular to the pentagon, when SPRs of very different energy or wavelength, and intensity can be tuned just by changing the light polarization.
\begin{figure}[htbp]
\begin{center} 
\includegraphics[width=  0.5\textwidth]{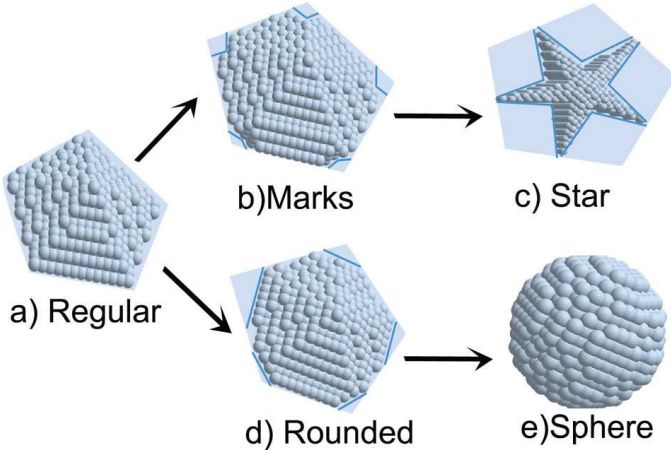}
\caption{Regular decahedron and  its truncated morphologies, and the sphere.} \label{fig8}
\end{center}
\end{figure}

When the size of the NP is in the range of 1~nm to 5~nm, the regular decahedron is never observed. The most common shapes are the truncated ones, the Marks decahedron and the round decahedron. The first structure was introduced by Marks~\cite{marks} and is remarkably stable and contains extra $\{111\}$ facets. In very clean growth conditions or with weak interactions with substrates, this is 
one of the predominant shapes for the discussed size interval.  An alternative way to describe the Marks decahedron is as a regular decahedron, which has truncations on its facets, as shown in Fig.~\ref{fig8}(b). When the truncation reaches a maximum value, a morphology with the shape of a star decahedron is formed, see Fig.~\ref{fig8}(c).  Another type of decahedral particle, which is often observed, corresponds to the round pentagonal particle. An example of these particles is shown in Fig.~\ref{fig8}(d). This kind of particle can be described as a truncated decahedron in which the truncation has a minimum possible value producing a contrast reduction in the borders. This type of particle is frequently formed when colloidal growth methods are used.~\cite{yacaman} Here, we discuss the optical response of the Marks decahedra with a truncation of $r=1/6$, and the maximum truncation of  $r=1/2$, which corresponds to the star decahedron, and we also discuss the rounded decahedron with a truncation of $r=1/8$.
\begin{figure}[htbp]
\begin{center} 
\includegraphics[width=  0.5\textwidth]{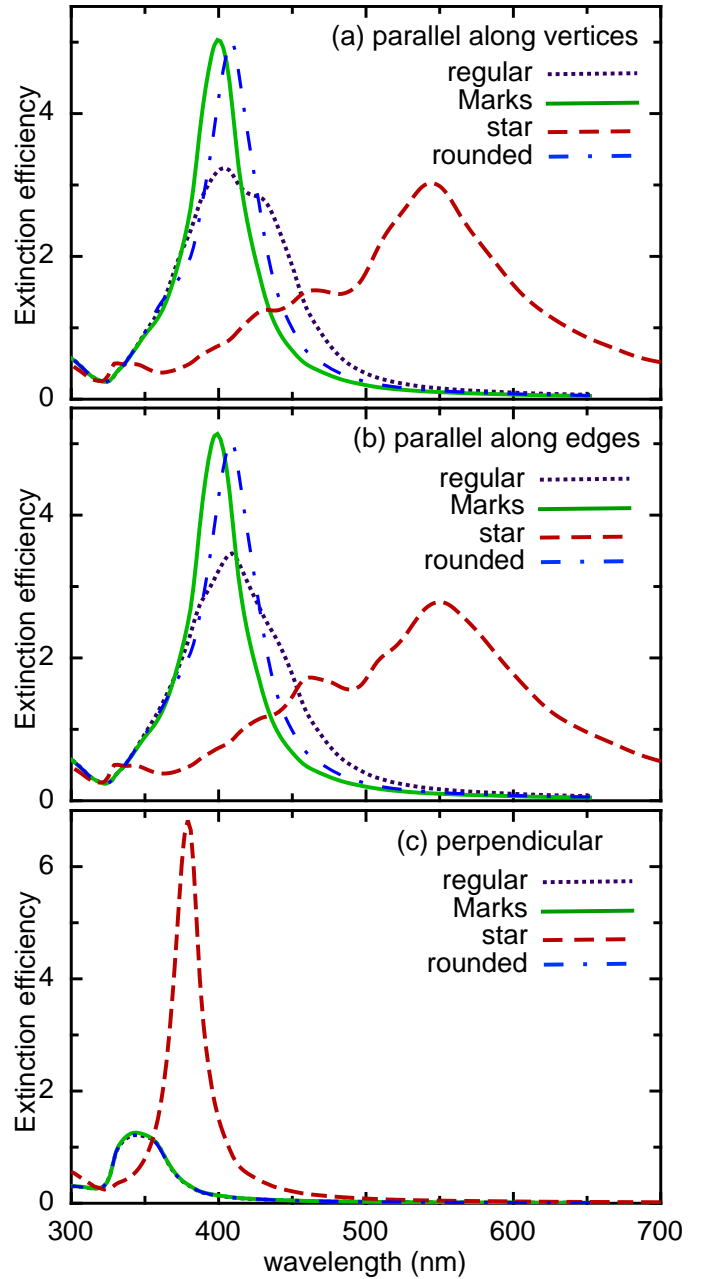}
\caption{Extinction efficiency as a function of the wavelength of the incident light of the regular decahedral nanoparticle and its truncated morphologies for different light polarizations. } \label{fig9}
\end{center}
\end{figure}

In Fig.~\ref{fig9}, the extinction efficiencies of the regular (dotted line), Marks (solid line), star (dashed line) and rounded (dashed-dotted line) decahedral nanoparticles for the three different polarizations are shown. Fig.~\ref{fig9}(a) corresponds to the case of parallel polarization along the vertices, and (b) to parallel polarization along the edges, while (c)  corresponds to the perpendicular one. We observe for the perpendicular polarization, Fig~\ref{fig9}(c), that the optical response of the regular decahedron does not change for small truncations, in both cases, the Marks and rounded decahedra. On the other hand, the response of the star decahedron is totally different since it shows a sharp resonance at 380~nm with a FWHM  of 20~nm, while for the other morphologies the maxima is at 343~nm, about seven times less intense, and has a FWHM of  45~nm. For both parallel polarization cases, all the spectra of the truncated decahedra show differences to respect to the regular one. For the rounded and Marks decahedra, the same effect is observed as in the case of truncated cubes. The main resonance is blue shifted, and becomes the most intense resonance after truncation, and also its FWHM decreases, as a result of the increment of the faces. On the other hand, the star decahedral shows the opposite behavior. In this case, the main resonance is red shifted to around 550~nm, and the spectra becomes very wide, since a lot of resonances are present. Comparing the star decahedron with the cube, we find some similarities, such as: (i) a large number of resonances located in a wide range of wavelengths; (ii) the main resonance is located at the right of the spectra. We also observe that these two morphologies present the sharpest vertices, such that, the charge distribution at the tips becomes very inhomogeneous,  leading to extreme field localization.~\cite{fuchs} 

Recently, single-crystal polyhedral NPs with uniform sizes have been synthesized.~\cite{tao} From TEM images, it was found that these polyhedral NPs exhibit defined facets with sharp edges and corners. Tao and collaborators~\cite{tao} reported that small silver NPs develop into cubes of 80~nm, and as their size grows, the NPs evolve from cubes to truncated cubes, cuboctahedra, and finally to octahedra of 300~nm. The different stages were characterized by TEM and optical spectroscopy, where the extinction spectra were measured. The optical spectra for cubes, cuboctahedra and octahedra show highly complex plasmon signatures as a result of their geometries. For such large NPs, the spectra show a red shift with increasing size as a consequence of the radiation effects. Taking into account the later, the overall behavior of the optical spectra as a function of the NPs geometry agrees well with our theoretical results. 

\section{Conclusions}

The influence of morphology on the optical properties of metal nanoparticles is studied theoretically using the discrete dipole approximation.  The location of the surface plasmon resonances of silver nanoparticles of different polyhedral shapes was obtained. We found that as the size truncation of the cubic nanoparticle becomes larger, the main resonance is blue shifted, overlapping secondary resonances, and therefore, increasing the  full width at half maximum of the main resonance. For decahedral particles, the truncation to Marks and rounded decahedra shows the same blue shift effect. However, the  full width at half maximum of the main resonance decreases, maybe because the secondary resonances no longer exist. It is also found that nanoparticles with fewer faces, like the star decahedron, show resonances in a wider range of wavelengths, perhaps because these geometrical shapes have  sharper vertices as compared to the others.  It is expected that this information would be useful to motivate the development of more complex nanostructures with tunable surface plasmon resonances.


\section*{Acknowledgments}  This work has been done with the partial financial support from CONACyT grant No.~44306-F and DGAPA-UNAM grant No.~IN101605.

\end{document}